\documentclass[letterpaper]{article} 
\usepackage{aaai23}  
\usepackage{times}  
\usepackage{helvet}  
\usepackage{courier}  
\usepackage[hyphens]{url}  
\usepackage{graphicx} 
\usepackage{natbib}  
\usepackage{caption} 
\usepackage{algorithm}
\usepackage{algorithmic}

\usepackage{multirow}
\usepackage{color,soul}
\usepackage{amssymb}
\usepackage{amsmath}
\usepackage{todonotes}
\usepackage{blindtext}
\usepackage{booktabs}
\usepackage{xcolor,colortbl}
\definecolor{LightGreen}{rgb}{0.8,1,0.89}
\definecolor{LightRed}{rgb}{1.0,0.8,0.7}
\definecolor{LightCyan}{rgb}{0.88,1,1}

\newcommand{\task}{\texttt{EXCLAIM}}
\newcommand{\dataset}{\texttt{ExHVV}}
\newcommand{\model}{\texttt{LUMEN}}
\newcommand{\geneval}{\texttt{gen-eval}}
\newcommand{\genevalc}{\texttt{gen-eval$_{\varsigma}$}}
\newcommand{\genevale}{\texttt{gen-eval$_{\vartheta}$}}

\usepackage{enumitem}
\newlist{tabenum}{enumerate}{1}
\setlist[tabenum]{wide=0pt, 
                  nosep, 
                  leftmargin= * ,
                  label*=\arabic*.,
                  after=\vspace{-\baselineskip},
                  before=\vspace{-0.6\baselineskip}}

\usepackage{newfloat}
\usepackage{listings}
\DeclareCaptionStyle{ruled}{labelfont=normalfont,labelsep=colon,strut=off} 
\floatstyle{ruled}
\newfloat{listing}{tb}{lst}{}
\floatname{listing}{Listing}
\title{{\em What do you MEME}? Generating Explanations\\for Visual Semantic Role Labelling in Memes}
\author{
    Shivam Sharma, \textsuperscript{\rm 1,\rm 4} Siddhant Agarwal, \textsuperscript{\rm 1} Tharun Suresh, \textsuperscript{\rm 1}\\ Preslav Nakov, \textsuperscript{\rm 2} Md. Shad Akhtar, \textsuperscript{\rm 1} Tanmoy Chakraborty\textsuperscript{\rm 3}}
\affiliations{
    \textsuperscript{\rm 1}Indraprastha Institute of Information Technology Delhi, India\\
    \textsuperscript{\rm 2}Mohamed bin Zayed University of Artifcial Intelligence, UAE\\
    \textsuperscript{\rm 3}Indian Institute of Technology Delhi, India\\
    \textsuperscript{\rm 4}Wipro AI Labs, India\\
    \{shivams, siddhant20247, tharun20119, shad.akhtar\}@iiitd.ac.in, preslav.nakov@mbzuai.ac.ae, tanchak@iitd.ac.in
}

\begin{document}

\maketitle

\begin{abstract}
Memes are powerful means for effective communication on social media. Their effortless amalgamation of viral visuals and compelling messages can have far-reaching implications with proper marketing. Previous research on memes has primarily focused on characterizing their affective spectrum and detecting whether the meme's message insinuates any intended harm, such as \textit{hate, offense, racism}, etc. However, memes often use abstraction, which can be elusive. Here, we introduce a novel task -- \task, generating explanations for visual semantic role labeling in memes. To this end, we curate \dataset, a novel dataset that offers natural language explanations of connotative roles for three types of entities -- \emph{heroes}, \emph{villains}, and \emph{victims}, encompassing 4,680 entities present in 3K memes. We also benchmark \dataset\ with several strong unimodal and multimodal baselines. Moreover, we posit \model, a novel multimodal, multi-task learning framework that endeavors to address \task\ optimally by jointly learning to predict the correct semantic roles and correspondingly to generate suitable natural language explanations. \model\ distinctly outperforms the best baseline across 18 standard natural language generation evaluation metrics. Our systematic evaluation and analyses demonstrate that characteristic multimodal cues required for adjudicating semantic roles are also helpful for generating suitable explanations. 
\end{abstract}

\section{Introduction}
\label{sec:intro}

In recent years, memes have become essential for communicating complex ideas and discussing societal issues. Besides being a developing phenomenon on the Internet, memetic visual recipes constantly mutate by adapting to the ever-evolving disparate social outlook, rendering them abstruse. Due to their design accessibility, memes are being increasingly disseminated to cause various kinds of harm \cite{pramanick-etal-2021-detecting}, including hate \cite{kiela_hateful_2020}, offense \cite{shang2021aomd}, trolling \cite{suryawanshi-chakravarthi-2021-findings}, etc. Recently, there have been developments towards automatically detecting such memes  \cite{Survey:2022:Harmful:Memes}. However, the limitations start materializing while contemplating aspects like multimodality and contextual dependency. 

For example, it is challenging to detect the multimodal narrative framing of social entities in memes \cite{sharma-etal-2022-disarm,sharma-etal-2022-findings}, and to model the inherent contextual reasoning.

\begin{figure}[t!]
    \centering
    \includegraphics[width=\columnwidth]{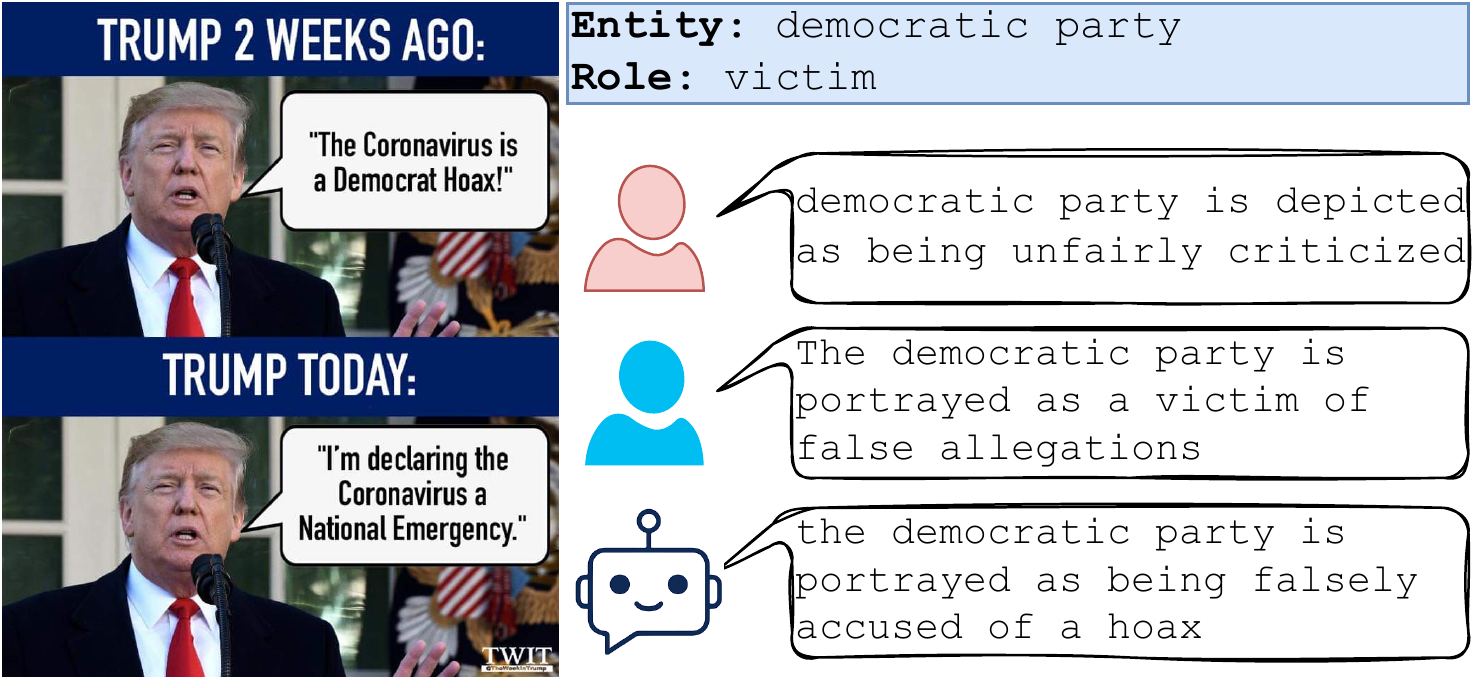}
    \caption{A novel task of \task. Given a meme, a referred entity and its semantic role, generate a natural language explanation that emulates human-like reasoning.}
    \label{fig:introfig}
\end{figure}

We attempt to leverage the features learned while detecting semantic role labels: \textit{hero, villain}, and {\em victim} for memes towards generating plausible natural language explanations. In doing so, we aim to contextualize memes towards critical social-media based use-cases like \textit{content moderation} and \textit{digital marketing}. Generated explanations can help contextualize role labeling and assist in the retrospective deduction for strategic insights.

Memes often implicitly ply sarcasm, satire, humor, and irony while vilifying, victimizing, or glorifying target entities. Generating explanations for such aspects is challenging due to the need for abstract reasoning and multimodal contextualization, as role labels and corresponding explanations are \textit{complementary} to each other. As an illustration, Figure~\ref{fig:introfig} depicts a meme in which \textit{Donald Trump} is portrayed as \textit{falsely} implicating the \textit{Democratic Party} for concocting the ruse of `Coronavirus', without \textit{explicitly} stating it. Therefore, \textit{Donald Trump} needs to be adjudicated as a \textit{villain}, based on the meme's semiotics involving his \textit{ironic stance-reversal} in the depicted scenario. Consequently, \textit{Democratic Party's} role as a \textit{victim} needs to be ascertained based upon the \textit{sarcasm} implied. \task\ asks for an explaining agent to factor in such intricacies while generating a suitable explanation for the aforementioned semantic roles.

The textual modality is well known to be instrumental towards detection of \textit{hateful memes} \cite{kiela_hateful_2020}, \textit{harmful memes, and their target types} \cite{pramanick-etal-2021-detecting,pramanick-etal-2021-momenta-multimodal}, \textit{meme emotion} \cite{sharma-etal-2020-semeval}, \textit{intra-modal incongruity} \cite{pan-etal-2020-modeling}, and \textit{semantic role labels} for \textit{multiple} affective connotations in memes \cite{sharma-etal-2022-findings}. On the other hand, a rare task exhibiting visual influence is \textit{detecting harmful entity} \cite{sharma-etal-2022-disarm}, which encompasses a \textit{single} affective connotation. Moreover, the image modality has been observed to under-perform in detecting meme emotions \cite{singh-etal-2020-lt3,ruiz-etal-2020-infotec}. Therefore, modality-influence typically varies across tasks. However, can something similar be generalized for the task complexity or modality-specific output configuration, like that solicited by \task?

\citet{bateman_2014} studied the text--image relation \cite{Barthes1978-BARI-84} graphically as a systematic network, and highlighted that text \textit{amplifies} the image, while an image \textit{inhibits} it. \citet{Osterroth} further deconstructed memes into three parts: (i) an initial verbal component influence \textit{generic} users, (ii) the visual part for \textit{experienced} users, and (iii) a conclusive second verbal part for the twists. This reinstates visual obscurity, emphasizing \textit{abstract reasoning} for critical memetic analysis. We aim to investigate contextual augmentation for \task\ via shared learning from multiple related objectives \citep{Lee2022}.


To summarize, we benchmark our proposed dataset, \dataset\, using several unimodal and multimodal baselines, typically involving uni/multi-modal encoder-decoder architectures. We further propose \model, a multimodal encoder-decoder framework that incorporates the entity-specific role-label information and explanation generation capability via multi-task learning. We compare the performances of these systems using multiple standard natural language generation (NLG) evaluation measures to assess their generation quality. We finally divulge \model’s limitations while highlighting the challenges posed towards addressing \task. Our contributions are summarized as follows:\footnote{The source code for this work can be found at: \url{https://github.com/LCS2-IIITD/LUMEN-Explaining-Memes}.}

\begin{enumerate}[leftmargin=*]
    \item \task: A novel task formulation, soliciting explanation generation for semantic role labelling in memes.
    \item \dataset: A multimodal dataset comprising natural language explanations accompanying the sets of memes, entities, and semantic roles.
    \item \model: A novel multimodal, multi-task learning framework that facilitates shared feature learning, from semantically related tasks.
    \item An extensive study of the explanation generation quality, across 18 standard NLG evaluation measures.  
\end{enumerate}

\section{Related Work}

This section briefly discusses relevant studies on meme analyses that primarily attempt to capture a meme's affective aspects, such as \textit{hostility} and \textit{emotions}, along with other noteworthy research on memes.

\paragraph{Meme Analysis.}
Several shared tasks have been organized lately, with a recent one on detecting the hero, the villain, and the victim entities in memes \cite{sharma-etal-2022-findings}. Others include troll meme classification \cite{suryawanshi-chakravarthi-2021-findings} and meme-emotion analysis via their sentiments, types and intensity prediction \cite{sharma-etal-2020-semeval}. Notably, hateful meme detection, introduced by \citet{kiela2020hateful} and later followed up by \citet{zhou2021multimodal}, garnered significant interest, with various solutions being proposed. A few of these efforts included fine-tuning Visual BERT \cite{li2019visualbert} and UNITER \cite{chen2020uniter}, along with using Detectron-based representations \cite{velioglu2020detecting,lippe2020multimodal} for hateful meme detection. On the other hand, there were systematic efforts involving unified and dual-stream encoders using Transformers \cite{muennighoff2020vilio,vaswani2017attention}, ViLBERT, VLP, UNITER \cite{sandulescu2020detecting,lu2019vilbert,zhou2020unified,chen2020uniter}, and LXMERT \cite{tan2019lxmert} for dual-stream ensembling. Besides these, other tasks addressed anti-semitism \cite{chandra2021subverting}, propaganda techniques \cite{dimitrov-etal-2021-detecting}, harmfulness \cite{pramanick-etal-2021-momenta-multimodal}, and harmful targeting in \cite{sharma-etal-2022-disarm} and detection of memes \cite{CURwebist20,memenonmeme}.

\paragraph{Visual Question Answering (VQA).}
Early prominent work on VQA with a framework encouraging \textit{open-ended} questions and candidate answers was done by \citet{VQA}. Since then, there have been multiple variations observed. \citet{VQA} classified the answers by jointly representing images and questions. 
Others followed by examining cross-modal interactions via attention types not restricted to {co/soft/hard-attention} mechanisms \cite{VQA_Parikh1,anderson2018bottom,eccvattention2018}, effectively learning the explicit correlations between question tokens and localized image regions. Notably, there was a series of attempts toward incorporating common-sense reasoning in
\cite{zellers2019recognition,wu2016ask,wu2017image,marino2019ok}. Many of these studies also leveraged information from external knowledge bases for addressing VQA tasks. 
General models like UpDn \cite{anderson2018bottom} and 
LXMERT \cite{tan2019lxmert} explicitly leverage non-linear transformations and Transformers for the VQA task, while others like LMH \cite{clark-etal-2019-dont} and SSL \cite{zhu2020prior}  addressed the critical language priors constraining the VQA performances, albeit with marginal enhancements.

\paragraph{Cross-Modal Association.}
Due to an increased influx of multimodal data, the cross-modal association has received significant attention lately. For cross-modal retrieval and vision-language pre-training, accurate measurement of cross-modal similarity is imperative. Traditional techniques primarily used concatenation of modalities, followed by self-attention towards learning cross-modal alignments \cite{cmr2016survey}. Following the object-centric approaches, \citet{zeng2021multi} and \citet{li2020oscar} proposed a multi-grained alignment approach, which captures the relation between visual concepts of multiple objects while simultaneously aligning them with text and additional meta-data. On the other hand, several methods also learned alignments between coarse-grained features of images and texts while disregarding object detection in their approaches \cite{huang2020pixel, kim2021vilt}. Later approaches attempted diverse methodologies, including cross-modal semantic learning from visuals and contrastive loss formulations \cite{yuan2021florence,jia2021scaling,clip2021radford}. 

Despite a wide coverage of cross-modal and meme-related applications in general, 
 there are still several fine-grained aspects of memes that are yet to be studied. 
Here, we attempt to address one such novel task: \task.

\begin{table}[t!]
\centering
\resizebox{0.7\columnwidth}{!}{%
\begin{tabular}{ccccccc}
\toprule
\multirow{2}{*}{\textbf{Split}} & \multicolumn{2}{c}{\textbf{U.S. Politics}} & \multicolumn{2}{c}{\textbf{Covid-19}} & \multicolumn{1}{c}{\multirow{2}{*}{\textbf{Test}}} & \multicolumn{1}{c}{\multirow{2}{*}{\textbf{Total}}} \\ \cline{2-3}\cline{4-5}

                                & \textbf{Train}        & \textbf{Val}       & \textbf{Train}     & \textbf{Val}     & \multicolumn{1}{l}{}                               & \multicolumn{1}{c}{}                                \\
\midrule
Villain                         & 1708                  & 217                & 654                & 78               & 347                                                & \textbf{3004}                                       \\
Victim                          & 531                   & 71                 & 357                & 47               & 104                                                & \textbf{1110}                                       \\
Hero                            & 276                   & 35                 & 185                & 20               & 50                                                 & \textbf{566}                                        \\ \hline
\textbf{Total}                  & \textbf{2515}         & \textbf{323}       & \textbf{1196}      & \textbf{145}     & \textbf{501}                                       & \textbf{4680} \\        \bottomrule                              
\end{tabular}%
}
\caption{Dataset summary, tabulating the entity-specific sample counts w.r.t. different \textit{domains} and \textit{splits}.}
\label{tab:dataset}
\end{table}

\section{\dataset: Dataset Curation and Annotation}


\paragraph{Dataset.}

Since \task\ requires a multimodal dataset with memes, semantic role labels for entities referred, and associated natural language explanations, we leverage HVVMemes. This dataset was recently released as part of a shared task at CONSTRAINT-2022 \cite{sharma-etal-2022-findings}. The original dataset, HVVMemes, was annotated considering the connotative labels for meme's entities by associating them with roles: \textit{hero, villain, victim} or \textit{others} across the domains of COVID-19 and US Politics. We augment HVVMemes with natural language explanations for the connotative labels provided for entities present in memes by re-annotating them and occasionally rectifying the role labels wherever obvious. Table \ref{tab:dataset} shows the summary of  \dataset. We consider only \textit{hero, villain} and \textit{victim}, as \textit{other} cases would be either inherently ambiguous or would solicit explanations that could be out-of-scope w.r.t. this work. This resulted in the increment of entity-specific sample count in \dataset\ for category \textit{hero} by 78, \textit{villain} by 616, and \textit{victim} by 159. Annotators are requested to explain why an entity could be a hero, villain, or victim. This resulted in a total of $4680$ labeled samples (c.f. Table \ref{tab:dataset}). The explanations reasonably varied across annotators, in terms of the \textit{vocabulary} and \text{length} owing to their different explanation styles. The subjectivity of the task is also highlighted in the annotations, as memes often lead to several valid interpretations.

\paragraph{Annotation Process.}
Two annotators were trained especially for obtaining explanations w.r.t. \task.\ They were prescribed standard annotation guidelines (c.f. Table \ref{tab:annotations}), which were drafted to assist them w.r.t. the task requirements and ambiguity resolution. The annotation guidelines explicitly emphasized the consideration of the meme author's viewpoint as a standardized reference. The annotators were also issued a list of verbs with certain use cases to maintain consistency across the annotations. Both the annotators were briefed w.r.t. annotation guidelines by a consolidator and were tasked with annotating an initial common set of 20 random memes with 125 entities for assessing and streamlining the annotation patterns. The test set was then annotated, considering the feedback from the consolidator. 
The remaining memes were then equally distributed amongst the annotators, with the consolidator adjudicating ambiguous cases. These aspects were accommodated in the final dataset to ensure consistent annotations.

\begin{figure*}[ht!]
    \centering
    \includegraphics[width=0.95\textwidth]{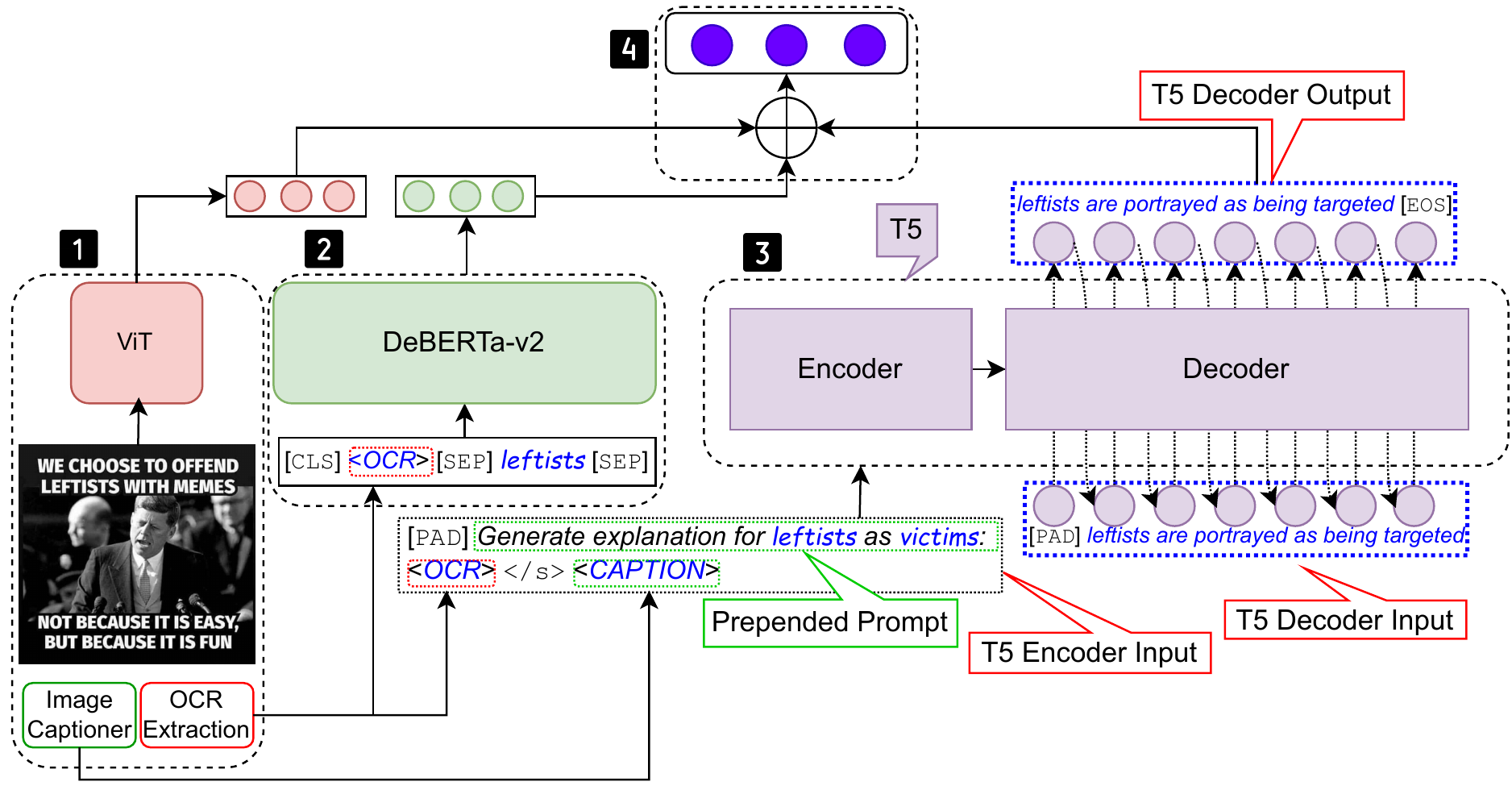}
    \caption{An illustration of \model's architecture, constituting modules: (1) Visual Recognition, (2) Entity Semantics, (3) Explanation Generation, and (4) Role-Label Prediction.}
    \label{fig:proposedarch}
    \vspace{-5mm}
\end{figure*}

The annotation quality was assessed via human evaluation by explicitly capturing the semantic validity and diversity of the explanations proposed. The questions posed in this assessment, along with the possible answer choices (and corresponding scores assigned) were:
\begin{itemize}
    \item Does the annotator provide a valid explanation? No/Partially/Yes (0/1/2), no label (-1)?
    \item Are the given pair of annotations diverse or not? No/Yes (0/1), invalid (no) annotation (-1)?
\end{itemize} 
The average normalized \textit{validity} and \textit{diversity} scores captured from this assessment were $0.81$ and $0.84$, respectively, for both the annotators.

\begin{table}[t!]
\centering
\resizebox{\columnwidth}{!}{%
\begin{tabular}{cp{8cm}}
\toprule
\textbf{S. No.} & \multicolumn{1}{c}{\textbf{Annotation Guidelines}} \\
\midrule
1    & Explanations should consider meme author's perspective.                            \\
2    & Explanations should emphasize meme's content only. \\
3    & Annotations should be narrated/written-            
\\
& \begin{minipage}{5in}
    \vskip 4pt
    \begin{itemize}
  \item Using reported speech for opinions.
  \item In the simple present form when stating facts.
  \end{itemize}
  \vskip 4pt
 \end{minipage}
                                                                                      \\
4    & Explanations should be lexically diverse.                                                                \\
5    & Cases labeled as \textit{others} should be skipped.                                                                                              \\
6    & Entity should constitute as explanation's primary subject.                                                                                            \\
7    & Obvious erroneous labels should be rectified.                                            \\
8    & Explanations may follow a standardized format: \small{\texttt{[Entity]~[Action/Abstract~Word/Phrases] [Description]}}.                              \\
9    & Each explanation should refer to a single entity.                                      \\
10   & Explanations may borrow facts from the OCR text.     \\
\bottomrule
\end{tabular}%
}
\caption{Prescribed guidelines for \dataset\ annotation.}
\label{tab:annotations}
\end{table}

\section{Methodology}

This section presents the details of \model, our proposed multimodal, multi-task learning framework to address \task\ (c.f. Fig. \ref{fig:proposedarch}). As exemplified while motivating \task\ in Section \ref{sec:intro}, the role labels and the corresponding explanations are likely complementary; we hypothesize that predicting one should help infer the other. \model\ comprises \textit{four} key components: (a) \textit{Visual Recognition}, (b) \textit{Entity Semantics}, (c) \textit{Explanation Generation}, and (d) \textit{Role-Label Prediction}. The latter three also contribute toward modeling the sub-tasks (ST) -- sequence classification (ST-1), explanation generation (ST-2), and role-label prediction (ST-3), as part of the \textit{multi-tasking} framework in \model.

The \textit{visual recognition} module contributes via OCR-based meme's text extraction, caption generation, and visual feature extraction. The \textit{entity semantics} module jointly encodes the meme's embedded text and the candidate entity while optimizing for \textit{sequence classification} task. On the other hand, the \textit{explanation generation} module firstly encodes a specially configured \textit{prompt}, along with the meme text and its captions, followed by optimizing over the autoregressive decoding of the encoded hidden representations. Finally, the outputs from the first three modules are exploited towards \textit{role-label prediction} task via their multimodal fusion. Effectively, \model\ jointly trains for sequence classification, role prediction, and explanation generation by optimizing the joint-loss formulation from the three corresponding objectives. We explicate each of these components in detail in the following subsections.

\subsection{Visual Recognition}
Since memes contain everything primarily as visuals, every input characteristic cue needs to be mined utilizing some visual processing mechanism first. To this end, we first extract the meme's embedded text and perform OCR-based extraction of embedded text from a given meme image using Google's OCR Vision API
We consider the meme's text entirely and discard any inherent line breaks while pre-processing. The meme's visual background in the form of imagery might not holistically capture the meme's intended message. However, it does provide a harmonizing comprehension for gaining a complete perspective on the meme's message. Evidence by \citet{blaier-etal-2021-caption} suggests that utilizing meme captions improves hateful meme identification results. Furthermore, additional cues such as the person, the location, and the entities present in the meme are helpful for downstream tasks. Thus, we use such ancillary information along with the OCR text. For image captioning, we make use of the recently-released OFA model \cite{wang2022ofa}. Finally, we encode the meme visuals using a ViT-based model and extract the pooled output from its last hidden layer, to obtain visual embedding $\mathbf{V}_{h_{i}}\in\mathbb{R}^{768}$.


\subsection{Entity Semantics}
As discussed previously, verbal content within the memes prominently constitutes towards characterizing their semantic undertones. Towards capturing the semantics from the textual content embedded within memes, we empirically designate the output from the last hidden layer of DeBERTa \cite{he2021deberta}, as our preferred choice. Besides demonstrating its superiority for a wide range of tasks, namely MNLI, SQuAD, RACE, etc., DeBERTa has demonstrated superior performance on detecting semantic role labels for entities in memes \cite{kun-etal-2022-logically}. We employ its \texttt{deberta-v2-xlarge} pre-trained variant ($\#$ layers: 24, H: 1536, and $\#$ parameters: 900M) for an objective similar to the latter, except for considering the ‘other’ category as a target semantic role within \task\ formulation. While using the last hidden layer output, we also fine-tune this setup toward the sequence classification objective. The inputs to this model are configured as a pair of sequences: \texttt{[CLS] A [SEP] B [SEP]}, wherein {\tt A} corresponds to OCR-extracted text and {\tt B} is the candidate entity. We fine-tune this model jointly within \model\ w.r.t. three target labels, \textit{hero, villain} and \textit{victim}, and obtain a mean-pooled representation for embedding \textit{entity-semantics}: $\mathbf{T}_{h_{i}}\in\mathbb{R}^{1536}$, along with a multi-class classification loss ($\mathcal{L}^{\text{SEQ}}$) for the given sequence,
\begin{small}
\begin{equation}
    \mathcal{L}^{\text{SEQ}} = -\sum_{c=1}^3y_{o,c}\log(p_{o,c})
\end{equation}
\end{small}

\subsection{Explanation Generation}
Generating natural language explanation conditioned on complex multimodal cues requires optimal modeling of entity-specific semantic role w.r.t. a given meme. To fully leverage the required information and compensate for the missing modality \cite{missingMa2021} induced due to obscure memetic visuals, we formulate the required encoder’s input as the combination of only text-based inputs, which inherently factors in the visual modality as well. To this end, we make use of  {T5}, a transformer-based encoder-decoder model, designed specifically to cater to tasks that can be reframed in \textit{text-to-text} format \cite{ColinT52020}. For encoder inputs, we consider meme’s \textit{OCR-extracted text} (source A) and \textit{caption} (source B), and configure them as \texttt{A [SEP] B [SEP]}. We also prepend every input with a \textit{task-specific prompt} as follows: 
\fbox{\begin{minipage}{0.97\columnwidth}
\texttt{Generate explanation for \textcolor{blue}{ENTITY} as \textcolor{blue}{ROLE}: \textcolor{blue}{OCR-TEXT} [SEP] \textcolor{blue}{CAPTION}}
\end{minipage}}
where, \textcolor{blue}{\texttt{BLUE}} items above are replaced by the corresponding values for each sample. This is also depicted in Figure \ref{fig:proposedarch}.

We use \texttt{t5-large} variant of T5, a 770 million parameter model checkpoint, originally pre-trained on a multi-task mixture of unsupervised and supervised tasks and was evaluated on a set of 24 tasks. We fine-tune it for the conditional generation objective along with the teacher-forcing strategy for training, and thereby obtain the T5-decoder’s mean-pooled last hidden layer representation ($\mathbf{E}_{h_{i}}\in\mathbb{R}^{1024}$) and a language modeling loss ($\mathcal{L}^{\text{EXP}}$), 
\begin{small}
\begin{equation}
    \mathcal{L}^{\text{EXP}} = -\log(p_{y_{t}}) = -\log(p(y_{t}|y_{< t}))
\end{equation}
\end{small}
\begin{table*}[t!]
\centering
\resizebox{0.8\textwidth}{!}{%
\begin{tabular}{clccccccc}
\toprule[1.5pt]
\multicolumn{1}{c}{\textbf{Modality}} & \multicolumn{1}{c}{\textbf{Model}} & \textbf{BLEU-1}$\blacktriangle$ & \textbf{BLEU-2}$\blacktriangle$ & \textbf{BLEU-3}$\blacktriangle$ & \textbf{BLEU-4}$\blacktriangle$ & \textbf{METEOR}$\blacktriangle$ & \textbf{ROUGE-L}$\blacktriangle$ & \textbf{CIDEr}$\blacktriangle$ \\ 
\midrule
\multicolumn{1}{c}{\multirow{5}{*}{UM}} & TXT-T5$^{\dagger}$ & \textbf{0.509} & 0.408 & 0.315 & 0.235 & 0.250 & 0.468 & \textbf{1.022} \\ 
\multicolumn{1}{c}{} & TXT-BERT.BERT & 0.498 & 0.396 & 0.313 & 0.230 & \textbf{0.254} & 0.468 & 0.878 \\ 
\multicolumn{1}{c}{} & TXT-BERT.GPT2 & 0.316 & 0.205 & 0.136 & 0.086 & 0.140 & 0.296 & 0.319 \\ 
\multicolumn{1}{c}{} & IMG-BEiT.GPT2 & 0.504 & \textbf{0.410} & 0.332 & \textbf{0.250} & 0.247 & 0.470 & 0.840 \\ 
\multicolumn{1}{c}{} & IMG-ViT.GPT2 & 0.489 & 0.389 & 0.309 & 0.230 & 0.238 & 0.456 & 0.816 \\ 
\midrule
\multicolumn{1}{c}{\multirow{7}{*}{MM}} & ViT.BERT.BERT & 0.498 & 0.404 & \textbf{0.404} & 0.239 & 0.253 & \textbf{0.473} & 0.890 \\ 
\multicolumn{1}{c}{} & ViT.BERT.GPT2 & 0.454 & 0.348 & 0.264 & 0.178 & 0.223 & 0.421 & 0.671 \\ 
\multicolumn{1}{c}{} & ViT.DeBERTa.GPT2 & 0.435 & 0.338 & 0.263 & 0.187 & 0.214 & 0.413 & 0.627 \\ 
\multicolumn{1}{c}{} & BEiT.BERT.GPT2 & 0.447 & 0.350 & 0.271 & 0.193 & 0.227 & 0.422 & 0.679 \\ 
\multicolumn{1}{c}{} & BEiT.DeBERTa.GPT2 & 0.445 & 0.350 & 0.274 & 0.198 & 0.222 & 0.427 & 0.706  \\ \cline{2-9} 
\multicolumn{1}{c}{} & \textbf{\model} & \textbf{0.578} & \textbf{0.485} & \textbf{0.399} & \textbf{0.313} & \textbf{0.294} & \textbf{0.530} & \textbf{1.380} \\ \midrule
\multicolumn{2}{c}{$\Delta_{\model\ - \dagger}$} & \textcolor{blue}{$\uparrow6.94\%$} & \textcolor{blue}{$\uparrow7.66\%$} & \textcolor{blue}{$\uparrow8.38\%$} & \textcolor{blue}{$\uparrow7.73\%$} & \textcolor{blue}{$\uparrow4.40\%$} & \textcolor{blue}{$\uparrow6.12\%$} & \textcolor{blue}{$\uparrow0.36$} \\ 
\bottomrule[1.5pt]
\end{tabular}%
}
\caption{Performance comparison for \task,\ using \geneval\ metrics across unimodal, multimodal and \model. Top two scores across the metrics are presented in bold; $\blacktriangle$ indicates higher scores are better; $\dagger$ represents the second best model.}
\label{tab:basecomp}
\end{table*}

\begin{figure*}[t!]
    \centering
    \includegraphics[width=\textwidth]{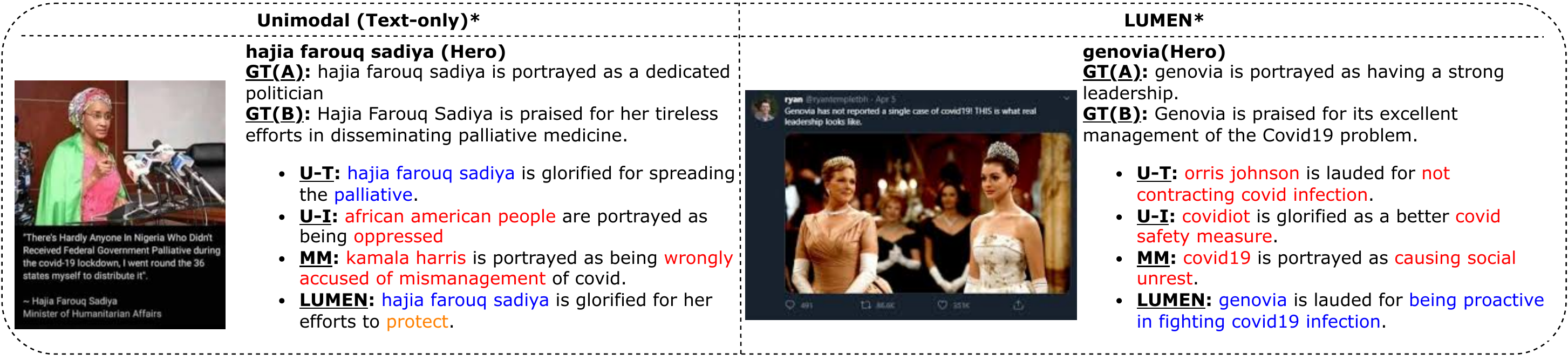}
    \caption{Illustration of explanation generation quality, for unimodal text (U-T), unimodal image (U-I), multimodal (MM) and \model. Left: U-T explains better; Right: \model\ explains better. Prediction scheme: \textcolor{blue}{Correct}, \textcolor{red}{Incorrect} and \textcolor{orange}{Partially-correct}.}
    \label{fig:discussion}
\end{figure*}

\subsection{Role-label Prediction}
Predicting semantic role labels for a given meme is another key sub-task that we aim to incorporate as part of \model\ design. Towards this, we fuse the representations gleaned via ViT-based visual features ($\mathbf{V}_{h_{i}}\in\mathbb{R}^{768}$), DeBERTa-based entity semantics ($\mathbf{T}_{h_{i}}\in\mathbb{R}^{1536}$) and T5-based representations obtained whilst decoding explanations ($\mathbf{E}_{h_{i}}\in\mathbb{R}^{1024}$). This not only takes into account modality-specific information for image and textual modalities via $\mathbf{V}_{h_{i}}$ and $\mathbf{T}_{h_{i}}$ respectively but also from the joint multimodal representation ($\mathbf{E}_{h_{i}}$), representing the interaction of visual description, textual context from meme and explanation-specific decoded hidden states. We first non-linearly project these representations individually into $512$ dimensional space before concatenating them. Finally, the fused representation is condensed into a $512$-sized embedding that represents the fused multimodal feature before eventually being linearly projected to $\mathbf{C}^{\text{OUT}}\in\mathbb{R}^{3}$ and utilized towards a 3-way multi-class classification. The corresponding cross-entropy loss ($\mathcal{L}^{\text{RP}}$) for semantic role prediction, culminated from this realization, is further incorporated into the overall multi-task learning objective.   
\begin{small}
\begin{equation}
    \mathcal{L}^{\text{RP}} = -\sum_{c=1}^3y_{o,c}\log(p_{o,c})
\end{equation}
\end{small}

Finally, we combine the loss terms obtained from entity-semantics ($\mathcal{L}^{\text{SEQ}}$), explanation generation ($\mathcal{L}^{\text{EXP}}$), and role prediction ($\mathcal{L}^{\text{RP}}$) modeling objectives. Optimizing using this joint-loss formulation facilitates leveraging the similarities of these three related tasks, leading to enriched feature learning in \model. The joint loss is configured as follows:
\begin{small}
\begin{equation}
    \mathcal{L}^{\text{LUMEN}} = \beta_{1}\mathcal{L}^{\text{SEQ}} + \beta_{2}\mathcal{L}^{\text{EXP}} + \beta_{3}\mathcal{L}^{\text{RP}}
\end{equation}
\end{small}
where $\beta$s are set as $\beta_{1}=0.2$, $\beta_{2}=0.5$, and $\beta_{3}=0.3$.

\section{Baselines}
We benchmark \dataset, using several multimodal and unimodal baselines. The baselines use transfer learning and are encoder-decoder models with unimodal and multimodal pre-trained encoders and a pre-trained text-based decoder fine-tuned on \dataset.
\paragraph{Unimodal Text-only.} T5 \cite{t5}, BERT \cite{DBLP:journals/corr/abs-1810-04805} and GPT2 \cite{radford2019language}.
\paragraph{Unimodal Image-only.} ViT \cite{Dosovitskiy2021AnII} and BEiT \cite{Bao2021BEiTBP}.

\paragraph{Multimodal.}
For the multimodal (text+image) systems, we employ several combinations of these vision encoders, text encoders, and text decoders as baseline models.
\begin{table}[t!]
\centering
\resizebox{\columnwidth}{!}{%
\begin{tabular}{lccccccc}
\toprule[1.5pt]
\multicolumn{1}{c}{\textbf{Model}} & \textbf{B-1}$\blacktriangle$ & \textbf{B-2}$\blacktriangle$ & \textbf{B-3}$\blacktriangle$ & \textbf{B-4}$\blacktriangle$ & \textbf{M}$\blacktriangle$ & \textbf{R-L}$\blacktriangle$ & \textbf{C}$\blacktriangle$ \\ 
\midrule
+ self-attend & 0.554 & 0.453 & 0.365 & 0.277 & 0.282 & 0.506 & 1.240 \\ 
\rowcolor{yellow}\textbf{\model} & \textbf{0.578} & \textbf{0.485} & \textbf{0.399} & \textbf{0.313} & \textbf{0.294} & \textbf{0.530} & \textbf{1.380} \\ 
- adafactor & 0.561 & 0.462 & 0.375 & 0.290 & 0.287 & 0.517 & 1.280 \\ 
- wtd loss & 0.559 & 0.461 & 0.372 & 0.283 & 0.285 & 0.504 & 1.313 \\ 
- captions & 0.531 & 0.433 & 0.347 & 0.265 & 0.276 & 0.486 & 1.198 \\ 
- T5 + GPT2 & 0.459 & 0.361 & 0.282 & 0.205 & 0.234 & 0.418 & 0.727 \\ 
- MTL & 0.435 & 0.338 & 0.263 & 0.187 & 0.214 & 0.413 & 0.627 \\ 
- deBERTa-v2 & 0.489 & 0.389 & 0.309 & 0.230 & 0.238 & 0.456 & 0.816 \\ 
- ViT & 0.509 & 0.408 & 0.315 & 0.235 & 0.250 & 0.468 & 1.022 \\ 
\bottomrule[1.5pt]
\end{tabular}%
}
\caption{Ablation results w.r.t. \model\ and its components, evaluated using \geneval. Top scores across the metrics are presented in bold; $\blacktriangle$ indicates \textit{higher} scores are better.}
\label{tab:ablation}
\vspace{-5mm}
\end{table}

\section{Experimental Details}
As part of our experimental setup, we first conduct multiple benchmarking experiments leveraging state-of-the-art systems as baselines and compare their performances with \model's. Since \task\ emulates the task family of natural language generation, we adopt an exhaustive set of standard evaluation metrics accordingly. Metrics used for comparing baseline performances: BLEU (B-[1,4]), METEOR (M), ROUGE-L (R-L), and CIDEr (C). Further, we discuss and compare explanations generated by unimodal (image/text-only), multimodal and \model\ systems. Next, we divulge ablation study-based performances towards establishing the component-wise relevance. Additionally, we evaluate best-performing models across modality configurations, using additional \textit{closeness-measuring} metrics: BERTScore, BP, chrF, GLEU, LASER, and RIBES, and \textit{error rates} based metrics: TER, WER, WER-D, WER-I, and WER-S. \footnote{We refer to the evaluation metrics for \textit{baseline} as \geneval; closeness-based comparison of top-performing systems set as \genevalc\ and error-rate-based metrics as \genevale.}
The experimental observations reported also capture the performance change induced by the proposed system, \model, w.r.t. the second best comparative baseline. Despite the metric-specific variations, the comparison is primarily made at the system level, and not at the individual metric level.

\subsection{Benchmarking \dataset}
Since our primary objective is to generate natural language explanations, we observe patterns contrary to what other related tasks conventionally exhibit, such as multi-class/label multimodal classification, wherein multimodal systems typically outperform unimodal ones. This performance behavior stems from the output requirement of \task,\ which is text generation. Also, this performance shift has varying implications regarding the explanation generation quality, which we will elaborate on later.

Unimodal text-only systems for BERT, GPT2, and T5-based decoders open with good median scores of $0.498$, $0.396$, $0.313$, and $0.230$ for BLEU-1/2/3/4, respectively, and  METEOR, ROUGE-L and CIDEr values $0.247$, $0.468$ and $0.840$ scores, respectively. This suggests not only better linguistic grounding but also objective completeness. Semantic alignment can be better adjudged by gleaning the generated explanations. In terms of \textit{uni-gram} overlap, the text-only T5-based system subtly outperforms the BEiT-based vision-only one by $0.5\%$, whereas, in terms of fluency (BLEU-2/3/4), the latter is better. The T5-based unimodal model yields a sizable gain of $18\%$ in terms of consensus-based CIDEr score, which propounds the dominance of unimodal text-only systems as being \textit{closer} to the ground-truth explanations (c.f. Table. \ref{tab:basecomp}). On the other hand, multimodal baselines (except ViT.BERT.BERT based model) exhibit a significant drop of $4\%$ median scores across the \geneval\ scores. This suggests a significant downgrade in the explanation generation quality that, to a reasonable extent, can also be characterized by the \geneval\ suite. Finally, \model,\ owing to its systematic, multimodal and multi-task regime, shows exceptional \geneval\ performance. 

As part of exploring MTL towards explanation generation for semantic role labeling of meme entities, our primary focus in this work is on explanation generation; therefore, we exhaustively evaluate it. As for the sequence generation task, we observe a steady decrease in validation loss from $0.1345$ to $0.1322$ over 15 epochs, with an average decrement rate of $1.5e-4$/epoch. For role-label prediction, we observe validation accuracies of up to $98\%$.

\subsection{\task: Analyzing Explanations}

The explanations by unimodal systems are sufficiently \textit{adequate} and \textit{fluent} while being brief and frequently accurate; decent median BLEU-1, BLEU-4, and ROUGE-L scores of 0.498, 0.230, and 0.468, respectively, indicate this. Their integrity is indicated by optimal median METEOR and CIDERr scores of 0.247 and 0.840, respectively. Text-based systems are often observed to get their short yet coherent explanations correct; for example, in Figure \ref{fig:discussion} (left), the T5-based generation correctly predicts the keyword - \textit{palliative} as the primary subject of the explanation, which other models struggle with. 
\model,\ on the other hand, shows its contribution to the \geneval\ suite by not only frequently generating semantically aligned explanations but also complete and factually relevant ones: see Figure \ref{fig:discussion} (right).

\subsection{Ablation Study}
This section presents the performance details (c.f. Table \ref{tab:ablation}), induced by sequentially removing critical components from \model.\ We begin by replacing the simple concatenation-based fusion in \model\ with popular multi-headed self-attention-based fusion and observe an average decline of $2\%$ across \geneval\ suite. This potentially indicates a self-aligning tendency of \model\ that operates using a simple concatenation of jointly projected visual, textual, and multimodal signals. Adafactor optimizer reflects better convergence and low memory footprint ($1\%$ drop on removal). Whereas \textit{wtd. loss} implies performing a weighted combination of the constituent loss terms ($0.5\%$ drop on removal) in multi-task learning (MTL) setup (c.f. Table \ref{tab:ablation}) in \model. Further, the textual captions from the meme visuals contribute crucial $2\%$ to the \geneval\ scores. Replacing the primary decoder, T5, with GPT2 brings down the average \geneval\ score massively from $0.390$ to $0.327$, reinstating the T5’s robustness for NLU tasks against that of GPT2. Multi-task learning setup demonstrates its utility with the performance reduction of $2\%$, when the system is evaluated with only text generation objective. Finally, removing deBERTa and ViT-based components prompts a reduction of $4\%$ and $3\%$, respectively, in the non-MTL configuration compared to the system under the MTL regime.    
\begin{table}[t!]
\centering
\resizebox{\columnwidth}{!}{%
\begin{tabular}{lcccccc}
\toprule[1.5pt]
\multicolumn{1}{c}{\textbf{Model}} &
\multicolumn{1}{c}{\textbf{BERTScore}$\blacktriangle$} & \multicolumn{1}{c}{\textbf{BP}$\blacktriangle$} & \multicolumn{1}{c}{\textbf{chrF}$\blacktriangle$} & \multicolumn{1}{c}{\textbf{GLEU}$\blacktriangle$} & \multicolumn{1}{c}{\textbf{LASER}$\blacktriangle$} & \multicolumn{1}{c}{\textbf{RIBES}$\blacktriangle$}         \\
\midrule
UM-TXT-T5$^{\dagger}$                  & \textbf{0.892} & 0.963          & \textbf{0.322} & 0.255          & 0.730           & \textbf{0.59} \\
UM-IMG-BEiT.T5             & 0.890           & \textbf{0.979} & 0.312          & 0.266          & 0.725          & 0.563         \\
MM-ViT.BERT.BERT          & 0.890           & 0.938          & 0.307          & \textbf{0.267} & \textbf{0.733} & 0.588         \\
LUMEN                      & \textbf{0.902} & \textbf{1.000}     & \textbf{0.368} & \textbf{0.280}  & \textbf{0.762} & \textbf{0.610} \\ \midrule
$\Delta_{\model-\dagger}(\%)$ & \textcolor{blue}{$\uparrow1.0$} & \textcolor{blue}{$\uparrow3.7$} & \textcolor{blue}{$\uparrow4.6$} & \textcolor{blue}{$\uparrow2.5$} & \textcolor{blue}{$\uparrow3.2$} & \textcolor{blue}{$\uparrow2.0$}\\
\bottomrule[1.5pt]
\end{tabular}%
}
\caption{Evaluation using \genevalc,\ that measures generated text's closeness with ground-truth. Top two scores across the metrics are presented in bold; $\blacktriangle$ indicates \textit{higher} scores are better; $\dagger$ represents the second best model.}
\label{tab:simeval}
\end{table}

\subsection{Extended Evaluation}
The $n$-gram-based metrics constituting \geneval\ overlook aspects like distributional semantics-based similarity, sentence length consideration, sub-word level overlap, multi-linguality, etc. Therefore, we further performed a comparative analysis of the best systems across modalities, including \model,\ using \genevalc,\ a suite of 6 additional evaluation metrics that represent explanation generation quality w.r.t. the ground-truth explanations (c.f. Table \ref{tab:simeval}). The trend observed earlier for \geneval\ is reflected again in \genevalc,\ with the models being in the decreasing order of \model,\ UM-TXT-T5, followed by UM-IMG-BEiT.GPT2 and MM-ViT.BERT.BERT, in terms of performance. The lead of $1.6\%$ and $1.1\%$ by the image-only model over the text-only one suggests the subtle closeness exhibited by image-only models.
\begin{table}[t!]
\centering
\resizebox{\columnwidth}{!}{%
\begin{tabular}{lccccc}
\toprule[1.5pt]
\multicolumn{1}{c}{Models} & \multicolumn{1}{c}{\textbf{TER}$\blacktriangledown$} & \multicolumn{1}{c}{\textbf{WER}$\blacktriangledown$} & \multicolumn{1}{c}{\textbf{WER-D}$\blacktriangledown$} & \multicolumn{1}{c}{\textbf{WER-I}$\blacktriangledown$} & \multicolumn{1}{c}{\textbf{WER-S}$\blacktriangledown$}  \\
\midrule
UM-TXT-T5$^{\dagger}$                  & \textbf{0.874} & 0.658          & 0.952          & 0.513          & \textbf{0.335}   \\
UM-IMG-BEiT.T5             & 0.9            & 0.652          & \textbf{0.828} & \textbf{0.497} & 0.341            \\
MM-ViT.BERT.BERT          & \textbf{0.861} & \textbf{0.634} & 0.952          & \textbf{0.331} & \textbf{0.332}   \\
\model                      & 0.907          & \textbf{0.634} & \textbf{0.519} & 0.633          & 0.361            \\
\midrule
$\Delta_{\model-\dagger}(\%)$ &  \textcolor{red}{$\uparrow3.3$} & \textcolor{blue}{$\downarrow2.4$} & \textcolor{blue}{$\downarrow43.3$} & \textcolor{red}{$\uparrow12.0$} & \textcolor{red}{$\uparrow2.6$}\\
\bottomrule[1.5pt]
\end{tabular}%
}
\caption{Error Evaluation using \genevale. Top two scores across metrics are presented in bold; $\blacktriangledown$ indicates \textit{lower} score are better; $\dagger$ represents the second best model.}
\label{tab:errorrate}
\end{table}

\section{Error Analysis}
Despite \model's stellar performance and consistent explanation generation quality, it is observed to fall short in exhibiting the required inductive biases acquired from \dataset. This is likely due to the inherent complexity posed by multimodal interactions, which either resist accurate prediction or induce additional noise in the pipeline. Another prominent limitation discerned from \model's explanations is the strong intra/cross-modal grounding, which in a few cases results in \model\ picking up some exact verbal or visual cues from the input memes and missing out on the correspondingly required semantic coherence. The impact of these challenges can be observed occasionally within the generated explanations. As for the \textit{error-rates} evaluated using \genevale,\ interestingly, \model\ yields increments of $3.3\%$, $12\%$, and $2.6\%$ on TER, WER-I, and WER-S, respectively. This not only implies the complexity involved in bridging the gap between the reference sentences and the candidate explanations but could also indicate the novelty, sufficiency, and creativity observed in \model’s explanations. This is corroborated by $2.4\%$ and $43\%$ deductions, \model\ induces in WER and WER-D scores, respectively. Also, \model-generated explanations exhibit greater diversity, with $45$ more unique non-stop words as against that for a unimodal text-only system.

\section{Conclusion}

We introduced \task, a novel task for generating natural language explanations for semantic role labels within memes. We first presented \dataset, our new multimodal dataset that curates natural language explanations suited for \task, and a benchmarking setup that encompasses multiple unimodal and multimodal baselines. Next, we proposed \model, an efficient and systematic multimodal, multi-task learning framework to address \task. 
The performance trends observed reflect clearly within the explanation generation quality for models from different modality configurations. Besides showcasing \model's adequacy and fluency, we also highlighted its inherent limitations, which constitute strong intra/cross-modal grounding w.r.t. the generated explanations 
and induced multimodal noise.
We hope that the insights from this work and the resources we release will nourish thought-provoking ideas to be explored in future work.   
\section*{Acknowledgement}
The work was supported by Wipro research grant.
\bibliography{aaai23}

\end{document}